\documentclass{article}
\usepackage{graphicx, natbib, enumerate,comment} 
\usepackage{titling}
\usepackage[a4paper, margin=2.5cm]{geometry}

\predate{}%
\postdate{}%

\usepackage{xcolor,amsmath}

\title{Viscosity of cooking oil: From kitchen to classroom}
\author{Endre Joachim Lerheim Mossige}
\date{}

\begin{document}

\maketitle

\begin{abstract}
    \noindent This paper presents a fun and relatable classroom experiment to measure the viscosity of cooking oil using household tools and ingredients like kitchen scales, mugs, and dried peas. The experiment requires no formal training in math or physics, and helps to build intuition about friction, Newton's laws of motion, and Archimedes' buoyancy law. 
\end{abstract}

\section*{Introduction}
The kitchen is a hub of curiosity and wonder. Why does water flow more easily than honey? Why does sour cream turn from solid to liquid when you stir it~\citep{nelson2022soft}? How do you make a perfectly boiled egg~\citep{di2025periodic}, and why are coffee stains darker near the edge~\citep{deegan1997capillary}? How are bubbles created in champagne~\citep{zenit2018fluid}? Shaken or stirred – or does it matter~\citep{trevithick1999shaken}? 

These familiar and sometimes surprising examples from our own kitchen cover a wide range of physics disciplines, from heat transfer (boiling an egg) and surface science (coffee stain) to soft matter physics (stirring sour cream) or thermodynamics (champagne bubbles)~\citep{rowat2014kitchen}. And at the core of it all lies fluid mechanics, the science of liquids and gases in motion, tying everything nicely together. In fact, the study of everything that flows in the kitchen, “kitchen flows”, also known as “culinary fluid mechanics”, is a rapidly growing research field~\citep{mathijssen2023culinary,fuller2022kitchen,herczynski2025coiling,kiyama2022morphology,park2025pour}.  

\begin{figure}[h]
    \centering
    \includegraphics[width=0.5\linewidth]{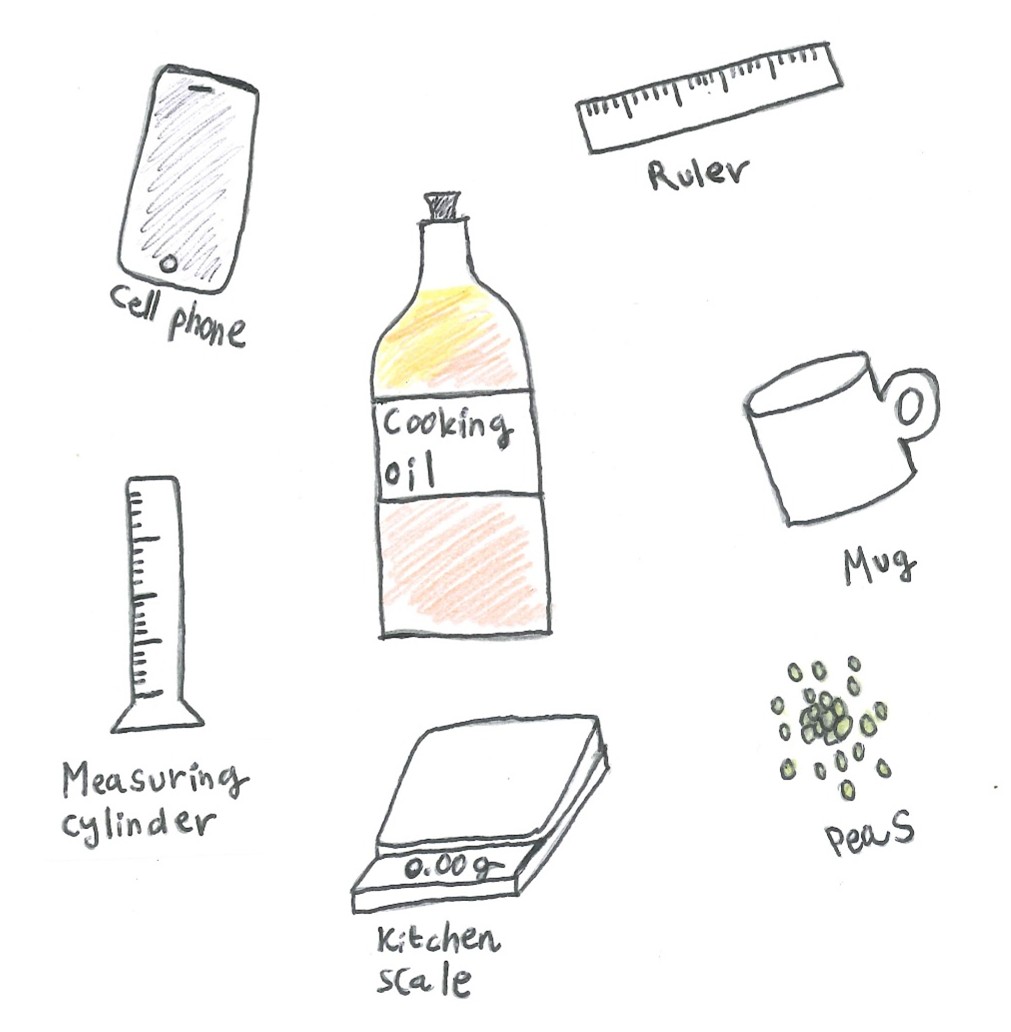}
    \caption{Experimental equipment used to measure the viscosity of cooking oil in the classroom experiment proposed in this paper: 5 kitchen scales, 5 measurement cylinders, 5 rulers, 1 liter of rapeseed oil, 10 small cups, 1 pack of frozen green peas. In addition, the students need to bring their cellphones for taking videos of peas sedimenting in oil.}
    \label{fig:ingredients}
\end{figure}


Kitchen flow experiments are both accessible and affordable and can be studied using household ingredients and simple tools with little or no formal training~\citep{mathijssen2023culinary}. And since we deal with them every day - when we make food, enjoy a delicious meal, or, inevitably, do the dishes - we have a natural intuition for them. As such, they are not just vehicles for tangible, curiosity driven research, but are indeed perfectly suited for fun and engaging classroom activities. Notably, Nigel Kaye created an online library describing over 70 cheap and easy-to-do classroom experiments aimed at students in an introductory fluid mechanics class ~\citep{nigelkaye}. The experiments relate to our daily encounters with fluid flows and includes kitchen flow activities such as blowing air out of a straw to learn about Newton's laws in a natural way~\citep{kaye2022overcoming}. 

The kitchen can also inspire new ways of learning about the material properties that control these flows. Two examples, both taken from the COVID era, are especially inspiring since they demonstrate how kitchen flows can be as an active learning strategy even in an online setting. In the first example, undergraduate students designed and performed tests in their own kitchen to probe the flow behavior of common household liquids like ketchup, yogurt, and buttercream frosting~\citep{hossain2022yourself}. Especially, the students performed tests to measure the "thickness", i.e. viscosity of these liquids, which is a measure of the internal friction in the fluid. In the second example, students learned about viscosity by making pancakes~\citep{ZenitPancakes}! If the batter spreads slowly, it is "thick" and has a high viscosity, while if it spreads more quickly, it is less "thick", or less viscous. Specifically, the students used their own cell phones to take videos to get the spreading rate, and fit these into a theoretical model~\citep{huppert1982propagation} to back-calculate the viscosity. Not only did the students get good data from this fun and simple experiment, but they also had a delicious breakfast~\citep{the}.

The experiments described above are easy to do, but their underlying physics is both complex and elaborate, and requires a solid background in physics or engineering. As such, they are best suited for university level students specializing in fluid mechanics or material science. To target a more general audience, this paper presents another tangible experiment with a much simpler physical underpinning, which is based on a familiar force balance from high school physics.  
It is inspired by a well established method for inferring viscosity, which involves dropping a steel ball with known size and density into a viscous fluid and measuring how fast it sediments. This method, which is called falling sphere viscometry~\citep{sutterby1973falling,brizard2005design}, is widely used in laboratories, and is a popular classroom exercise.

\begin{figure}[h]
    \centering
    \includegraphics[width=0.7\linewidth]{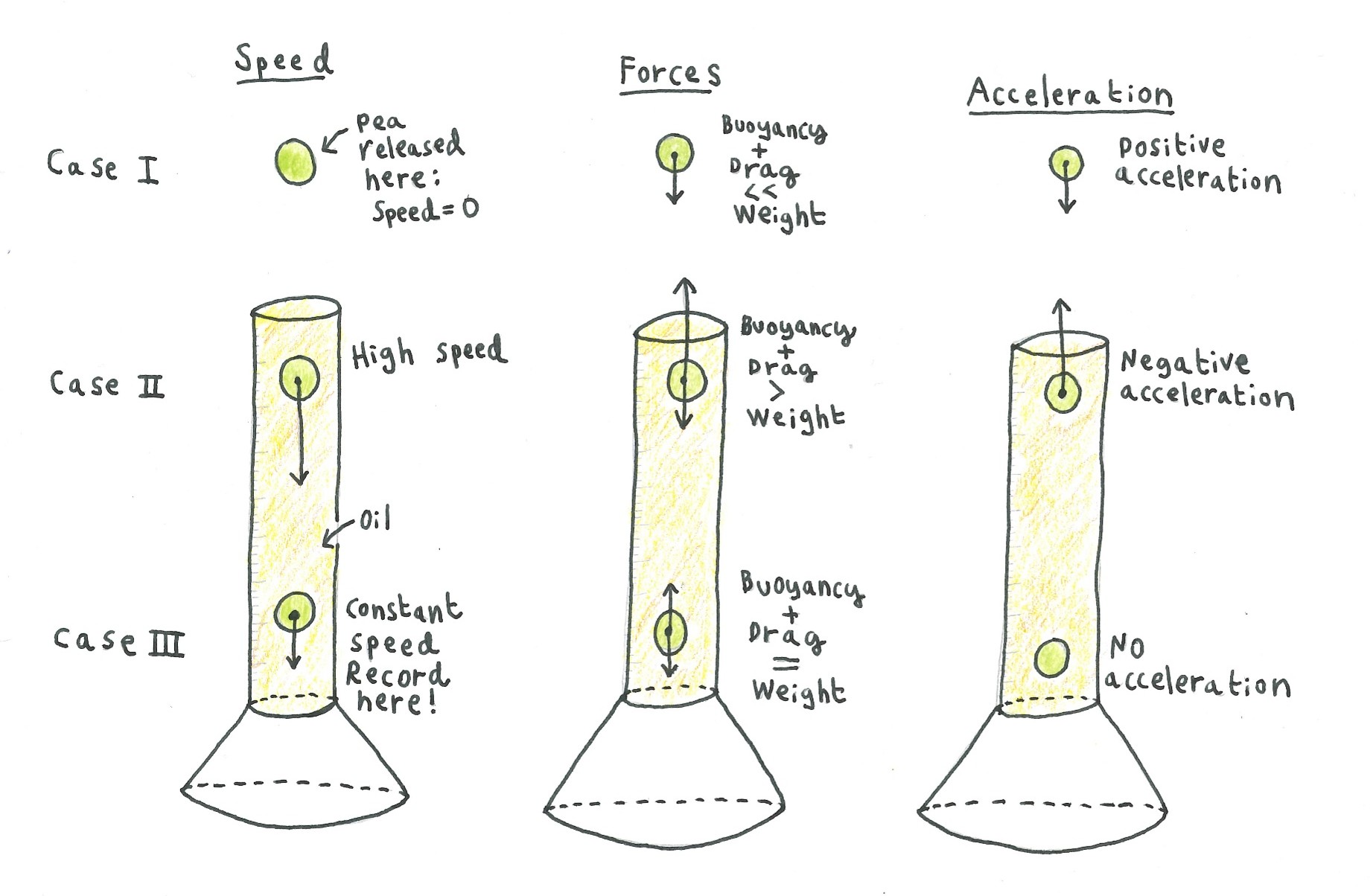}
    \caption{The figure shows velocity, force, and acceleration diagrams for three special cases relevant to our viscosity measurement. Case I: The pea falls freely in air, with zero buoyancy, and with negligible drag. Case II: The pea enters the oil phase at high speed, giving rise to a high viscous drag force, which leads to a negative acceleration. Case III: The weight force balances the buoyancy plus viscous drag force so that the pea falls at a constant speed.}
    \label{fig:blackboard}
\end{figure}

This paper shows how the falling sphere experiment can be made even more fun and relatable by using tools and ingredients found in most households, see Figures \ref{fig:ingredients} and \ref{fig:blackboard}. Instead of using uniform steel balls falling through idealized liquids like castor or silicone oil, the students use dried peas sedimenting in rapeseed oil from the supermarket. And instead of using laboratory scales and calipers to accurately measure the weight and size of individual peas, the students use simple kitchen scales and rulers to obtain average quantities. This approach naturally poses questions like “How many peas do I have to measure to get good data”, and “How many times do I have to repeat the measurement?”. As such, this engaging classroom experiment not only helps the students develop intuition about friction, buoyancy and Newton’s laws, which are central ingredients in physics education, but also fosters critical thinking and ownership by actively taking part in their own learning. 

This paper describes a 90-minute lecture which was given to students with no formal training in math or physics as part of the general physics class "Physics for the public" at the University of Oslo \citep{FFFF}. The first part of the lecture, described in Chapter 1, explores the force balance on falling objects, and explains how we can infer the viscosity of cooking oil by considering peas falling at constant speed. The second part of the lecture, described in Chapter 2, lays out a five-step experimental recipe used to conceptualize theoretical topics such as Newton's laws of motion, and Archimedes' buoyancy law. Chapter 3 presents the experimental data and experiences from the general physics class. In particular, this chapter examines why peas fall more slowly in narrow cylinders, and how this observation can teach us about friction. Chapter 4 compares the experimental data with literature values and discusses potential routes for improvement, and is aimed at trained physicists and engineers specializing in fluid mechanics. Chapter 5 concludes the paper.




\section{Theory: Force balance on a falling sphere}
Consider a round object falling through a viscous liquid, for example a pea falling through cooking oil, see Figure \ref{fig:blackboard}. The speed and acceleration of the falling object depends on the relative magnitude of three forces, namely the weight force, which points downwards, the buoyant force, which points upwards, and the viscous drag force, which also points upwards. This third force is a resistance force required to displace nearby liquid, and it increases with both the speed and the size of the falling object, as well as the "thickness", or viscosity, of the fluid. 

Newton's second law states that the sum of forces ($\sum \mathrm{Forces}$) on an object is equal to its mass $m$ times its acceleration $a$:

\begin{equation}
    \sum \mathrm{Forces = Weight - Buoyancy - Drag}=m \cdot a. 
    \label{eq:sum_of_forces}
\end{equation}
Three special cases are of interest to us, and for all of them, the falling pea helps to build intuition, see Figure 2. The first case concerns a pea released in air, falling faster and faster. Here, the only force contributing to its motion is the weight force since the viscous drag force and the buoyancy force can be neglected due to the low viscosity and density of air. The second case concerns a pea immediately after entering the oil phase, where it slows down rapidly due to both buoyant and viscous drag forces. As the pea continues to fall through the oil, the buoyant and viscous forces counteract its downward motion, until they exactly balance the weight force. In this third and most important case (because it allows us to compute the viscosity), the pea sediments at constant speed with no acceleration. 



Using mathematical symbols, we write the weight force as $\mathrm{Weight} = m_{pea}g$, where $g$ is the acceleration constant. The buoyant force is given by Archimedes' law: $\mathrm{Buoyancy} = m_{oil}g = \rho_{oil} V_{pea} g$, where $m_{oil}$ is the mass of the oil displaced by the pea, $\rho_{oil}$ is the density of the oil (measured in grams per milliliters), and $V_{pea}$ is the volume of the pea. Finally, the drag force on a small sphere falling slowly in a viscous liquid like oil depends on the viscosity, $\mu_{oil}$, the sedimentation speed, $U_{pea}$, the diameter, $d_{pea}$, and the shape constant $3\pi$, and is given by the celebrated Stokes' drag law~\citep{stokes1851effect}:  

\begin{equation}
    \mathrm{Drag} = 3\pi \mu_{oil} U_{pea}d_{pea}.
    \label{eq:drag}
\end{equation}

\noindent Using symbols, the sum of forces on a pea falling slowly through viscous oil is:
\begin{equation}
    \sum \mathrm{Forces}=m_{pea}g - \rho_{oil} V_{pea}g - 3\pi \mu_{oil} U_{pea}d_{pea} = m_{pea} \cdot a_{pea}.
    \label{eq:sum_of_forces_symbols}
\end{equation}

We consider peas falling through cooking oil at constant speed (Case III) since this allows us to infer the viscosity. Here, the sum of forces is zero:

\begin{equation}
    \sum \mathrm{Forces} = m_{pea}g - \rho_{oil} V_{pea}g - 3\pi \mu_{oil} U_{pea}d_{pea} = 0. 
    \label{eq:sum_of_forces_pea}
\end{equation}

\noindent Solving for the viscosity of oil yields

    \begin{equation}
    \mu_{oil} = \frac{\left(m_{pea} -\rho_{oil} V_{pea}\right)g}{3 \pi d_{pea} U_{pea}}.
    \label{eq:viscosity_of_oil}
\end{equation}

In the classroom experiment below, the students measure the unknown properties above using the kitchen tools in Figure \ref{fig:ingredients} and plug them into Eq. \eqref{eq:viscosity_of_oil} to calculate the viscosity, $\mu_{oil}$. 

\section{Classroom experiment: Getting your hands wet}

\subsection{Prepping for the experiment}
Before the classroom experiment begins, let the students play with the rapeseed oil (touching it, pouring it, without using the measurement equipment) and based on this playing-with-your-food-experiment, the students make a prediction of how viscous and how dense viscosity is compared to water. In the classroom experiment described below they will test their hypotheses. The teacher may also ask the students how they would go about designing an experiment using the household equipment in Figure 1 to measure the missing quantities on the right-hand-side of Eq.\eqref{eq:viscosity_of_oil}. This can be done as a think-pair-share exercise, where the students first think for themselves, then pair with their partner to discuss their strategy, and finally share their thoughts with the rest of the class. 

\subsection{Experimental recipe}
\begin{figure}[h]
    \centering
    \includegraphics[width=0.8\linewidth]{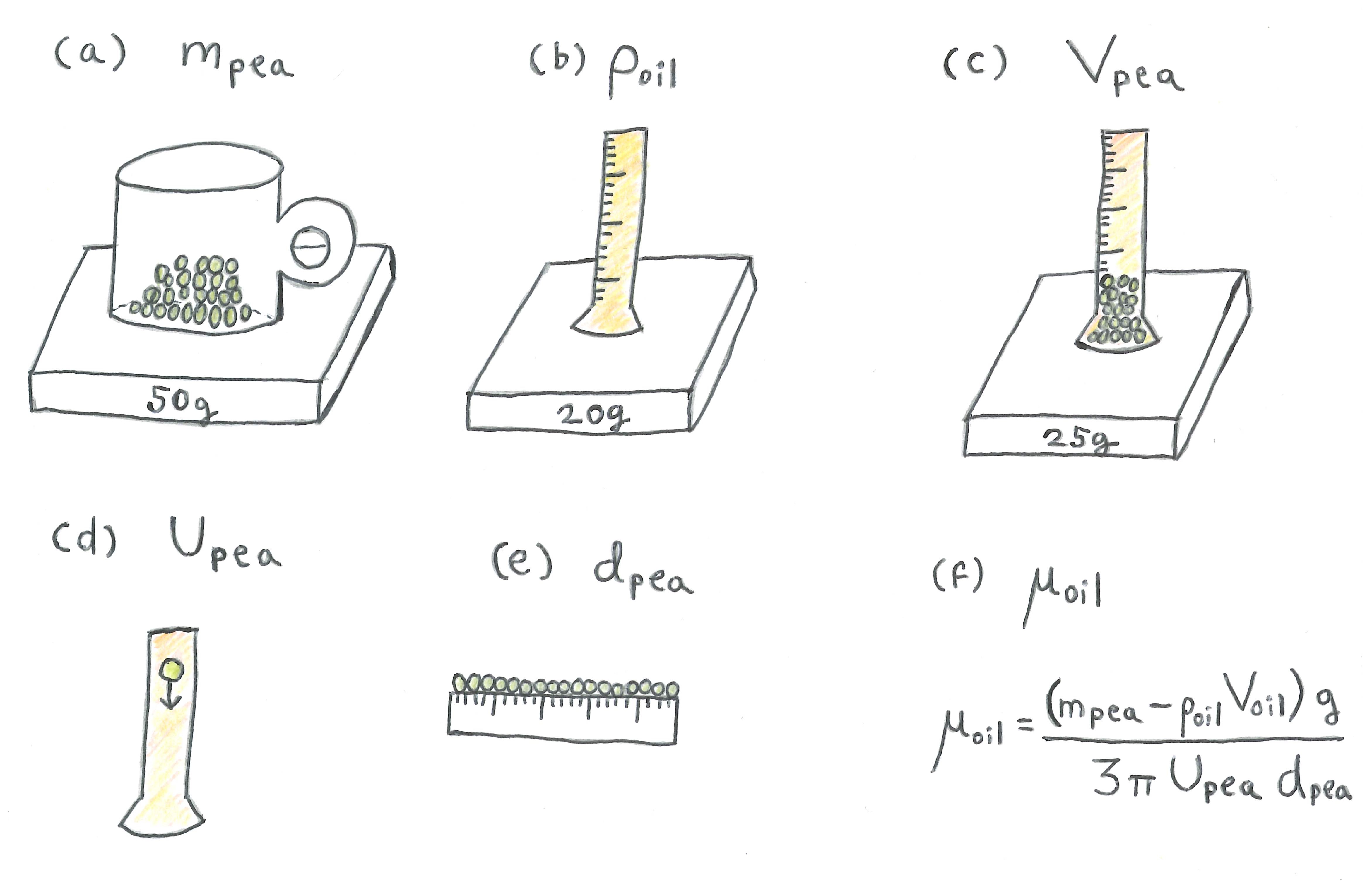}
    \caption{Experimental recipe used to infer the viscosity of cooking oil.}
    \label{fig:recipe}
\end{figure}

To back-calculate the viscosity from Eq. \eqref{eq:viscosity_of_oil}, the students can either use their own experimental protocol, or they can use the following recipe, see Figure \ref{fig:recipe}: 

\begin{enumerate}
    \item $m_{pea}$: Measure the mass of a typical pea using the kitchen scale. 
    \item $\rho_{oil}$: Measure the density of oil by weighing the fluid and reading off the volume using the measuring cylinder. 
    \item $V_{pea}$: Measure the volume displaced by a typical pea. 
    \item $U_{pea}$: Measure the sedimentation speed: Record with your phone!
    \item $d_{pea}$: Measure the diameter of a typical pea using the ruler. Alternatively, calculate $d_{pea}$ from $V_{pea}$ using the formula for the volume of a sphere: $V_{pea}=\pi d_{pea}^3/3$. Which one is more accurate?  
    \item $\mu_{oil}$: Calculate the viscosity from Eq. \ref{eq:viscosity_of_oil}. Evaluate: Is it close to what you predicted?
\end{enumerate}

\begin{figure}[t]
    \centering
    \includegraphics[width=0.8\linewidth]{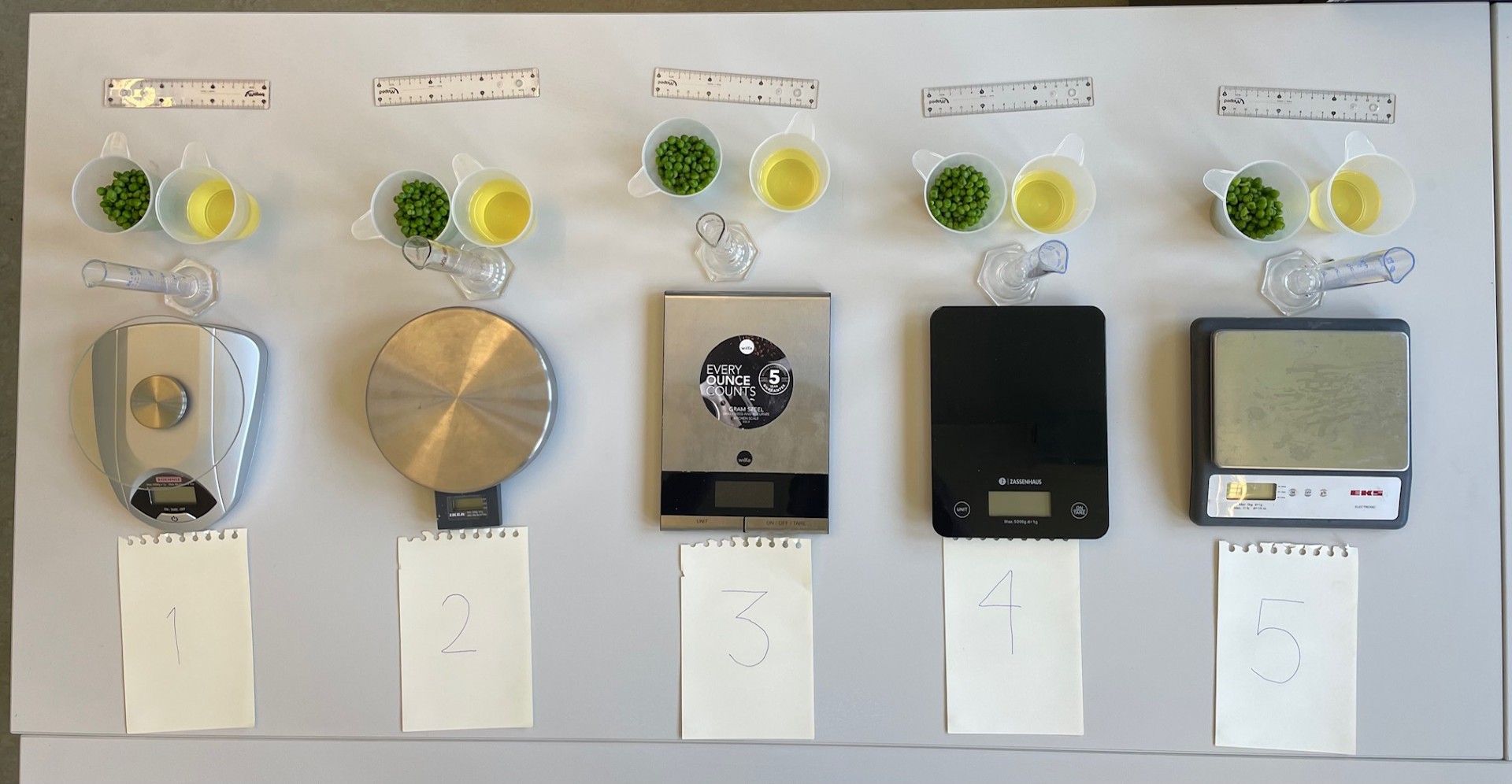}
    \caption{Experimental equipment used in the classroom experiment: 5 kitchen scales, 5 measurement cylinders, 5 rulers, 1 liter of rapeseed oil, 10 small cups,  1 pack of dried peas. In addition, the students bring their cellphones for taking videos of peas sedimenting in oil.}
    \label{fig:prepping}
\end{figure}

Once the teacher has presented the experimental recipe, he or she divides the students into groups, and distributes the necessary kitchen tools, see Figure \ref{fig:prepping}. Preferably, the measuring cylinders given to the different groups vary in size, and in Chapter 3: "Experimental results" we demonstrate how using cylinders with diameters ranging from 11 to 27 mm in the general physics class was used to learn about fluid friction. Finally, to keep track of the students' progress, it is a good idea to make a table on the blackboard where they fill in their measurement values. This also helps them to self-evaluate their own progress, keep track of time, and to correct for outliers by peeking at the other groups' data.

\section{Results from the classroom experiment}
\subsection{Experiences from the classroom experiment}
With the simple set of instructions presented in the last chapter, the students in the general physics class worked independently and systematically. Being hands-off as teacher allowed the students to fully concentrate on the experiments and build team spirit, in turn giving them ownership to their own learning. Yet, to develop important critical thinking skills, which is essential to solve physics problems~\citep{trusikova2020critical}, the teacher motivated the students to consider the following: How many peas did you have to measure to get good data? Is the value of $\rho_{oil}$ and $\mu_{oil}$ close to what you predicted? Did you expect $\rho_{oil}$ to be smaller than the density of water, and why? Which of the experimental steps (a-e) produced the largest error, and why? Is the density of peas, $\rho_{pea}=m_{pea}/V_{pea}$, larger than the density of oil (as it should be)? In the general physics class, these questions informed a lively discussion between the groups, and they could also be addressed a lab report or homework assignment after the lecture. 

\subsection{Experimental data from the classroom experiment}
Table \ref{table:results_classroom_nmbu2026} shows the experimental data obtained by the different groups participating in the classroom experiment. We note that while the data for $m_{pea}$, $\rho_{oil}$, $V_{pea}$, $\rho_{pea}$, and $d_{pea}$ differ only slightly between the groups, the data for the sedimentation speed, $U_{pea}$, vary much more, spanning from 1.90 to 7.91 $cm/s$, yielding a ratio of 4.2. 

The large spread in the speed data is partly due to differences in measuring method between the different groups. For example, some groups used a stopwatch to measure the transit time of falling peas, while other groups took videos. And while some groups released the peas immediately above the oil-air surface, other groups released them within the oil phase. One group even decided to wet the peas in oil before they dropped them and noticed that pre-wetted peas fall faster than initially dry peas. This observation initiated a fruitful discussion between the group members, who speculated that the reduced sedimentation speed for initially dry peas is due to small air bubbles being trapped on their surface, giving them additional buoyancy. 

However, while the different experimental protocols accounted for some of the spread in the data, the main reason is that the different groups used measuring cylinders with vastly different diameters, ranging from 11 to 27 $mm$. In a think-pair-share exercise, the students reflected on this finding, and they rightly concluded that the walls in confined geometries like measuring cylinders slow the fluid around falling objects (for example peas), giving them additional drag, which in turn causes them to fall slower \citep{guyon2015physical}. After this enlightening discussion, the students were sent home to prepare for the final exam, and the Appendix lists a few suitable problems. 

The rest of the paper was not part of the lecture, and is better suited for fluid mechanics students at the university level.

\begin{table}[t]
\begin{center}
\begin{tabular}{ |l|c|c|c|c|c|c|c|c|c| } 
 \hline
 Parameter & $D_{cylinder}$ &$m_{pea}$ & $\rho_{oil}$ & $V_{pea}$ & $\rho_{pea}$ &  $U_{pea}$ & $d_{pea}$ &
 $\mu_{oil}$ &
 $\mu_{oil}/\mu_{water}$\\
 Unit & $mm$ & $g$ & $g/ml$ & $ml$ & $g/ml$ &  $cm/s$ & $mm$ & $mPa\cdot s$ & $-$\\
 \hline
Group 1 &11& 0.26 & 0.85 & 0.19 & 1.34 & 1.90 & 6.00 & 865 & 865\\ 
Group 2 &18& 0.24 & 0.95 & 0.18 & 1.32 & 5.54 & 6.42 & 196 & 196\\ 
Group 3 & 20& 0.22 & 0.84 & 0.20 & 1.08 & 7.01 & 7.00 & 102 & 102\\ 
Group 4 & 20 & 0.24 & 0.86 & 0.18 & 1.36 & 7.32 & 6.45 & 195 & 195\\ 
Group 5 & 27 & 0.22 & 0.82 & 0.19 & 1.14 & 7.91 & 6.61 & 124 & 124\\ 
 \hline
\end{tabular}
\caption{Measurement results from experiments performed by students in a general physics class. Here, $m_{pea}$ is the mass of a typical pea, $\rho_{oil}$ is the density of the oil, $V_{pea}$ is the volume a typical pea displaces, $\rho_{pea}$ is the density of peas ($\rho_{pea}=m_{pea}/V_{pea}$; should be higher than $\rho_{oil}$ since the pea sinks. Not needed in the calculation of viscosity, but merely used to control the  measurement of $m_{pea}$, $V_{pea}$, and $\rho_{oil}$), $U_{pea}$ is the sedimentation speed, $d_{pea}$ is the diameter of peas, $\mu_{oil}$ is the dynamic viscosity of the oil, and $\mu_{oil}/\mu_{water}$ is the dynamic viscosity of oil relative to water at standard classroom conditions (1.002 $mPa\cdot s$ at 1 atm pressure and 20 $^o C$).}
\end{center}
\label{table:results_classroom_nmbu2026}
\end{table}

\section{Diving deeper: Comparison with literature values and routes for improvement}

\subsection{Correcting for wall effects} The additional drag due to wall effects explained above can be compensated for using a correction factor, $K_p$, which depends on the relative size of the falling object to the size of the geometry. This ratio is called the fill ratio, $d_{pea}/D_{cylinder}$ in our case, where it spans from 0.55 to 0.24, see Table 2. \citet{haberman1958motion} computed the wall correction factor $K_p$ for fill ratios up to 0.8:

\begin{equation}
    K_p =
\frac{
1 - 0.75857 \left(\frac{d}{D}\right)^5
}{
1 - 2.1050 \left(\frac{d}{D}\right)
+ 2.0865 \left(\frac{d}{D}\right)^3
- 1.7068 \left(\frac{d}{D}\right)^5
+ 0.72603 \left(\frac{d}{D}\right)^6
}.
\label{eq:Kp}
\end{equation}
Substituting $d_{pea}$ for $d$ and $D_{cylinder}$ for $D$ in Eq. \eqref{eq:Kp} above yields the $K_p$-values given in Table 2. In the same table, we divide the viscosity values obtained by the students, $\mu_{oil}$, by $K_p$ to compensate for wall induced drag to yield a more accurate viscosity via $\mu_{oil}^{\prime} = \mu_{oil} / K_p$.  

\begin{table}[b]
\begin{center}
\begin{tabular}{ |l|c|c|c|c|c|c| }
\hline
Parameter & Unit & Group 1 & Group 2 & Group 3 & Group 4 & Group 5 \\ \hline
$D_{cylinder}$ & mm & 11 & 18 & 20 & 20 & 27 \\
$d_{pea}/D_{cylinder}$ & - & 0.55 & 0.36 & 0.35 & 0.32 & 0.24 \\
$K_p$ & - & 7.58 & 2.97 & 2.89 & 2.58 & 1.94 \\
$\mu_{oil}$ & $mPa\cdot s$ & 865 & 196 & 102 & 195 & 124 \\ 
\hline
$\mu_{oil}^{\prime} = \mu_{oil} / K_p$ & $mPa\cdot s$ & 127 & 66 & 35 & 75 & 64 \\
\hline
\end{tabular}
\caption{Wall correction factors, $K_p$, as function of the fill ratio $d_{pea}/D_{cylinder}$ for the different groups, where $d_{pea}$ is the pea diameter and $D_{cylinder}$ is the diameter of the measuring cylinder used in the velocity measurement. The $K_p$-values are used to correct for an additional drag force due to additional friction from the cylinder walls in the viscosity measurement via $\mu_{oil}^{\prime} = \mu_{oil} / K_p$, where $\mu_{oil}$ is the viscosity value without wall-correction as measured by the students.}
\end{center}
\label{table:Kp}
\end{table}

\subsection{Comparing with literature values} The viscosity of oil drops down sharply with temperature, as anyone who has poured oil into a frying pan can appreciate. As such, to facilitate a direct comparison between our viscosity data with data reported previously, we need to make sure the measurements were done at the same temperature. Using a laboratory plate-plate rheometer, \citet{brodin2025interface} measured the viscosity of rapeseed oil to be 73 m$Pa \cdot s$ at 20$^\circ C$, which matches the temperature in the classroom experiments.

\begin{figure}[h]
    \centering
    \includegraphics[width=0.8\linewidth]{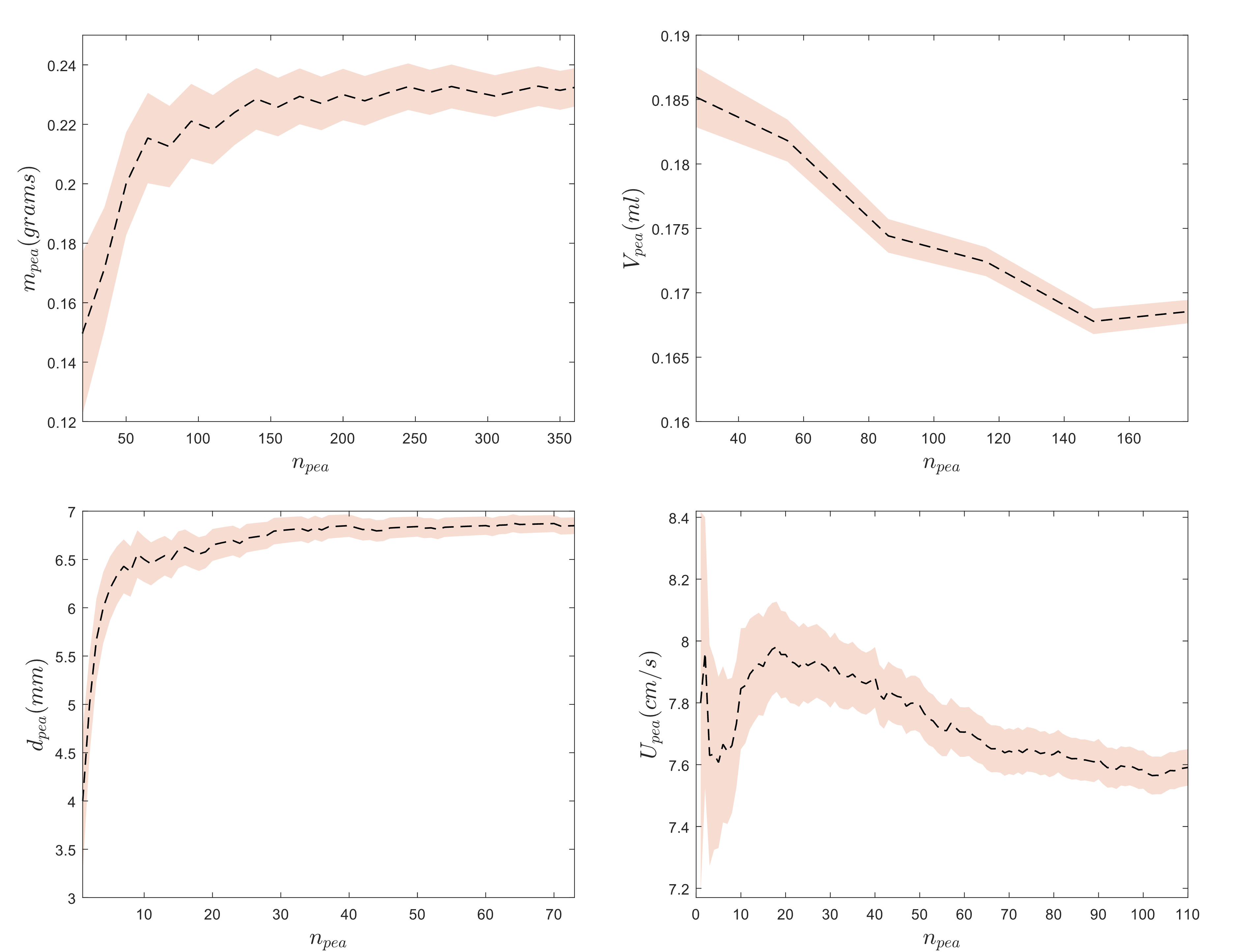}
    \caption{Mean values of mass, volume, diameter and speed of peas ($m_{pea}$, $V_{pea}$, $d_{pea}$, and $U_{pea}$) as function the number of peas used to get the mean. The shaded regions indicate the so-called sample mean error, which gives the error of the mean value. A 100 mL measuring cylinder with an inner diameter of 27 mm was used in the measurements.}
    \label{fig:homeexperiments}
\end{figure}

Our viscosity values, listed in Table 2, differ from the literature value of viscosity, and this should be expected since our data were obtained using simple tools from the kitchen, while \citet{brodin2025interface} used a professional rheometer to measure the viscosity. Our experiment is far from perfect, and as pointed out by the students, it is prone to error as the peas may not fall in a straight line, and as they may generate bubbles as they are dropped in oil.  
In addition, the peas do not all have the same size, which means that the measurements are sensitive to the number of peas used to obtain the average value of mass, volume, diameter and sedimentation speed ($m_{pea}$, $V_{pea}$, $d_{pea}$, and $U_{pea}$). Figure \ref{fig:homeexperiments} shows how the mean values of these four quantities depend on the number of peas used in the measurement to compute the mean, and the shaded regions indicate the so-called sample mean error, which is a measure of the expected error of the mean value (see the Appendix for details). When we are dealing with average quantities (as in this case and so often in physics experiments), the mean value should not vary significantly with successive experiments. For the pea speed measurement (bottom right panel in Figure \ref{fig:homeexperiments}) for example, this requires as much as 100 or more repeats. In the classroom experiment, the maximum number peas used to measure the sedimentation speed done by any of the groups was 18 (and the minimum number was 3), likely giving an over-prediction.

\subsection{Routes for improvement: Accounting for inertia}
As well as being prone to experimental error, the simple classroom experiment presented in this paper violates the creeping (slow) flow condition of Stokes' drag law in Eq. \eqref{eq:drag}, which states that inertial forces should be much smaller than viscous forces. Their ratio is given by the so-called Reynolds number, which in our case is 

\begin{equation}
    Re = \frac{\rho_{oil}U_{pea}d_{pea}}{\mu_{oil}^{\prime}}.
    \label{eq:Reynoldsnumber}
\end{equation}

So, to obey Stokes' creeping flow condition, $Re$ must be smaller than one, $Re < 1$, while in our experiment, $Re$ spans from 0.8, 5.1, 11.7, 5.4, and 6.7 for group 1 thru 5. Stokes' drag can be recast in terms of the cross-sectional area of the falling particle, in our case $A_p =\pi d_{pea}^2/4$, the dynamic pressure, $p_{dyn}=1/2\rho_{oil}U_{pea}$, and the dimensionless drag coefficient, $c_D$:

\begin{equation}
    \mathrm{Drag} = \frac{1}{2}c_DA_pp_{dyn},
    \label{eq:DragWith_cD}
\end{equation}

\noindent where 

\begin{equation}
    c_D = \frac{24}{Re}
    \label{eq:StokesDragcD}
\end{equation}

in the Stokes' drag regime for which $Re<1$. To account for flow inertia, \citet{oseen1910stokes} proposed the following improved formula for the drag coefficient:
    \begin{equation}
    c_{D} = \frac{24}{\text{Re}}\left( 1+\frac{3}{16} \text{Re} \right),
    \label{eq:OseenDrag}
    \end{equation}
The first term in Eq.\eqref{eq:OseenDrag} is the Stokes drag near the particle, while the second term accounts for inertial effects far from the particle. Oseen's formula agrees fairly well with experiments up to $Re \approx 10$~\citep{dey2019terminal}, covering all but one of the Reynolds numbers in our experiments. As such, combining \eqref{eq:OseenDrag} with \eqref{eq:DragWith_cD} to calculate the drag in our experiments would bring us closer to the true viscosity value. 

\section{Conclusion}

This paper presents a simple classroom experiment using kitchen tools to measure the viscosity of rapeseed oil, with the primary goal of developing student intuition for Newton's second law of motion by bridging theory with observation.    

\section*{Acknowledgments}
The author thanks Are Raklev at the University of Oslo for organizing the general physics course "FYS1050 – Physics for the public" and for the kind invitation to guest lecture, and Vegard Nilsen at the Norwegian University of Life Sciences for the kind invitation to give a preliminary version of this lecture to students in a fluid mechanics class. The author also thanks Leiv Rønneberg for clarifying inputs on the statistical analysis and he acknowledges feedback on the manuscript from Dag Dysthe. Finally, the author thanks the wider fluid mechanics community for showing interest in "kitchen flow" experiments as an active learning strategy.    

\clearpage

\section*{Appendix}
\label{sec:appendix}

\subsection*{Standard deviation and the sample mean error}

The standard deviation is given by:
\begin{equation}
    S = \sqrt{\frac{\Sigma\left(\hat{x}-x \right)^2}{n-1}},
    \label{eq:stdev}
\end{equation}
where $\hat{x}$ is the measured value for the n-th measurement, $x$ is the mean value, and $n$ is the number of measurements in the ensemble.

The sample mean error gives the error of the mean value, and is given by

\begin{equation}
    S_{LR} = \frac{S}{\sqrt{n}}.
\end{equation}

Since $S_{LR} \sim 1/\sqrt{n}$ it should decrease with successive measurements $n$. 

\subsection*{Proposed Exam Questions in the general physics course}

Correct answers are underlined.

\subsubsection*{Problem 1}
The forces acting on a round pea submerged in a liquid (e.g., cooking oil) are the buoyant force, which acts upward, and the gravitational force, which acts downward. The forces are equal in magnitude. What will happen to the pea?

\begin{itemize}
\item It will move upward
\item It will move downward
\item It is impossible to answer based on the information given in the problem statement
\item \underline{It will not move}
\end{itemize}

\subsubsection*{Problem 2}
A small oil droplet (diameter: 1\,mm) rises with constant velocity in water. Is the gravitational force acting on the droplet greater or smaller than the buoyant force?

\begin{itemize}
\item Greater
\item \underline{Smaller}
\item Equal
\item It is not possible to answer based on the information given in the problem statement
\end{itemize}

\subsubsection*{Problem 3}
A pea is released from a height into a container with cooking oil. Due to frictional forces (drag force), the speed decreases so that it falls slower and slower for a while, before eventually reaching a constant velocity. During the deceleration phase, i.e., while it is falling slower and slower (negative acceleration), the forces acting on the pea are:

\begin{itemize}
\item Equal
\item The forces acting downward are greater than the forces acting upward
\item \underline{The forces acting upward are greater than the forces acting downward}
\item It is not possible to answer based on the information given in the problem statement
\end{itemize}

\subsubsection*{Problem 4}
When a pea falls in cooking oil, a frictional force (drag force) acts on it. What causes this frictional force?

\begin{itemize}
\item \underline{The pea has to push away viscous fluid in front of it}
\item The pea has to push away viscous fluid behind it
\item The pea absorbs viscous fluid
\item The pea repels viscous fluid
\end{itemize}

\clearpage

\bibliography{viscosity}

\end{document}